\documentclass[prd,twocolumn,floatfix,preprintnumbers,letterpaper]{revtex4}
\usepackage{graphics}
\usepackage{epsfig}
\usepackage{amsmath}

\begin{document}

\title{Purely kinetic $k$-essence as unified dark matter}
\author{Robert J. Scherrer}
\affiliation{Department of Physics and Astronomy, Vanderbilt University,
Nashville, TN  ~~37235}

\begin{abstract}
We examine $k$-essence models in which the Lagrangian $p$ is a function only of the derivatives of
a scalar field $\phi$ and does not depend explicitly on $\phi$.
The evolution of $\phi$
for an arbitrary functional form for $p$ can be given in terms of an exact analytic solution. For quite general
conditions on the functional form of $p$, such models can evolve
to a state characterized by a density $\rho$ scaling with the scale factor $a$ as $\rho = \rho_0 +
\rho_1 (a/a_0)^{-3}$, but with a sound speed $c_s^2 \ll 1$ at all times.  Such models can
serve as a unified model for dark matter and dark energy, while avoiding the problems of the
generalized Chaplygin gas models, which are due to a non-negligible sound speed in these models.
A dark energy component with $c_s \ll 1$ serves to suppress cosmic microwave background
fluctuations on large
angular scales.

\end{abstract}

\maketitle

\section{Introduction}

The universe appears to consist of approximately 25\% dark matter, which
clusters and drives the formation of large-scale structure in the universe, and
70\% dark
energy, which drives the late-time acceleration of the universe (see Ref.
\cite{Sahni} for a recent review, and references therein).  Since the nature of
neither component is known with certainty, it is reasonable to ask whether a
simpler model is possible, in which a single component acts as both dark matter
and dark energy.  In this letter, we show that a certain class of scalar
field models with non-standard kinetic terms, dubbed $k$-essence, can serve
as such a unified model for dark matter and dark energy.

The idea of $k$-essence was first introduced
as a possible model for inflation \cite{Arm1,Garriga}.  Later, it was noted that $k$-essence can
also yield interesting models for the dark energy
\cite{Chiba1,Arm2,Arm3,Chiba2,Chimento1,Chimento2}.  It
is possible to construct a
particularly interesting class of such models in which the $k$-essence energy density tracks the
radiation energy density during the radiation-dominated era, but then evolves toward a
constant-density dark energy component during the matter-dominated era \cite{Arm2,Arm3}.  In
this class of models, the coincidence problem (i.e., why we live in the particular era during
which the dark matter and dark energy densities are roughly equal) is resolved by linking the
onset of dark energy domination to the epoch of equal matter and radiation.  (Note that the
plausibility of such models was challenged by Malquarti et al. \cite{Malquarti1}, who argued
that the basin of attraction for models exhibiting the desired behavior is actually quite
small).  A second interesting characteristic of $k$-essence models is that they can produce a
dark energy component with a sound speed smaller than the speed of light.  Such models can be
observationally distinguished from standard scalar field quintessence models with a
canonical kinetic term (for which $c_s = 1$), and may provide a mechanism to
supress cosmic microwave background (CMB) fluctuations on large angular scales
\cite{Erickson,DeDeo,Bean}.

In this paper, we reexamine a particularly simple class of $k$-essence
models, in which the Lagrangian contains only a kinetic factor, i.e., a function of the derivatives
of the scalar field, and does not depend explicitly on the field itself.  Equivalently, these
models can be considered to have a constant potential term.  Such models were, in fact, the first
ones investigated
in the context of inflation \cite{Arm1}.  In this context, they successfully yield
exponential inflation, but suffer from a ``graceful exit" problem.  Here, we are interested in
such models as a model for dark energy and dark matter.  In the next section, we examine such
models in the generic case, and show that, for quite general assumptions on the functional
dependence of the Lagrangian on the scalar field derivatives, these models naturally produce a
density which scales like the sum of a non-relativistic dust component with equation of state
$w \equiv p/\rho = 0$ and a cosmological-constant-like component $(w = -1)$.  The other
distinguishing characteristic of these models is that they generically produce a low sound speed,
$c_s^2 \ll 1$, allowing the ``dust" component to cluster as dark matter.  Our results are
summarized in Section III.

\section{Purely Kinetic $k$-essence}
\subsection{General Solution}
In general, $k$-essence can be defined as
any scalar field with non-canonical kinetic terms, but in practice such models are usually taken
to have a Lagrangian of the form:
\begin{equation}
\label{p}
p = V(\phi)F(X),
\end{equation}
where $\phi$ is the scalar field, and $X$ is
\begin{equation}
\label{grad}
X = \frac{1}{2} \nabla_\mu \phi \nabla^\mu \phi.
\end{equation}
The pressure in these models is simply given by equation (\ref{p}), while the energy density is
\begin{equation}
\label{rho}
\rho = V(\phi)[2X F_X - F],
\end{equation}
where $F_X \equiv dF/dX$.
Therefore, the equation of state parameter, $w \equiv p/\rho$, is just
\begin{equation}
\label{w}
w = \frac{F}{2X F_X - F}.
\end{equation}

In defining the sound speed, we follow the convention of Garriga and Mukhanov \cite{Garriga}, who
argued that the relevant quantity for the growth of density perturbations is
\begin{equation}
\label{cs}
c_s^2 = \frac{(\partial p /\partial X)}{(\partial \rho/\partial X)} = 
\frac{F_X}{F_X + 2X F_{XX}},
\end{equation}
with $F_{XX} \equiv d^2 F/dX^2$.

We will work in a flat Robertson-Walker metric, in which the equation for the $k$-essence field take the
form:
\begin{equation}
\label{motion1}
(F_X + 2X F_{XX})\ddot \phi + 3H F_X \dot \phi + (2XF_X - F)\frac{V_\phi}{V} = 0,
\end{equation}
where $V_\phi \equiv dV/d\phi$,
$H$ is the Hubble parameter, given by
\begin{equation}
\label{H}
H^2 = (\rho/3),
\end{equation}
and we take $8 \pi G =1$ throughout.  In equation (\ref{H}),
$\rho$ is the total density, including the contributions of
the background radiation and matter, as well as the $k$-essence
component.

A variety of functional forms for $F(X)$ and $V(\phi)$ appearing in equation
(\ref{p}) have been considered.  For example, Armendariz-Picon et al. \cite{Arm2,Arm3}
concentrated on $V(\phi)$ of the form $V(\phi) =  1/\phi^2$, and they then found
various forms for $F(X)$ for which the $k$-essence field tracks the radiation
density in the radiation-dominated era, but acts like a cosmological constant in
the matter-dominated era.  Tachyon fields correspond to the choice \cite{Chiba2}
$F(X) = \sqrt{1-2X}$, with a variety of potentials $V(\phi)$ having been
investigated.  Here we consider one of the simplest possible $k$-essence models:
a purely kinetic model in which
\begin{equation}
p =  F(X),
\end{equation}
so that $V$ is set to a constant.  As
noted above, this simple class of models was first investigated in Ref.
\cite{Arm1} as a model for inflation, and, of course, it represents a special
limiting case of many of the more general models examined in Refs.
\cite{Chiba1,Arm2,Arm3,Chiba2,Chimento1,Chimento2}.

Note that the equation of state, $w$, and the sound speed, $c_s^2$, do not
depend explicitly on $V(\phi)$ in any case; their dependence on the choice of the
potential arises entirely through the effect of $V$ on the
evolution of
$\phi$ (equation \ref{motion1}).  Taking $V$ to be a constant, we get
a new equation for $\phi$:
\begin{equation}
\label{motion2}
(F_X + 2X F_{XX})\ddot \phi + 3H F_X \dot \phi = 0,
\end{equation}
which can be rewritten entirely in terms of $X$:
\begin{equation}
\label{motion3}
(F_X + 2X F_{XX})\dot X + 6H F_X X = 0.
\end{equation}
Changing the independent variable from the time $t$ to the scale factor $a$, gives
\begin{equation}
(F_X + 2X F_{XX})a \frac{dX}{da} + 6 F_X X = 0,
\end{equation}
which can be integrated exactly, for arbitrary $F$, to give the solution:
\begin{equation}
\label{solution}
X F_X^2 = ka^{-6},
\end{equation}
where $k$ is a constant of integration.  This solution was previously derived, in a slightly
different form, by Chimento \cite{Chimento2}.
Given any form for $F(X)$, equation (\ref{solution}) gives the evolution of $X$ as a function of $a$,
and this solution can be substituted into equations (\ref{p}),(\ref{rho}),(\ref{w}), and (\ref{cs}) to determine the evolution of all of the
physical quantities of interest.  In the next section, we consider a particularly interesting example.

\subsection{Evolution near an extremum in F}

Suppose that $F(X)$ is a function with
a local minimum or maximum at some value $X=X_0$, so that
$F_X(X_0) = 0$.    For example, Chiba et al. \cite{Chiba1} examined Lagrangians of the form
\begin{equation}
\label{Chibaeq}
p = K(\phi) X + L(\phi) X^2.
\end{equation}
With $K$ and $L$ simply taken to be constants, $K = K_0$, $L = L_0$, this Lagrangian can be thought of as a
canonical kinetic term plus a non-standard correction.  This Lagrangian has an extremum at
$X = - K_0/2L_0$.  We will consider, however, the general case of an arbitrary function $F$ with an extremum at
$X_0$.

Then equation
(\ref{motion3}) has the obvious solution $X = X_0$, i.e., a constant value of
$X$.  This result is neither new nor remarkable; it was first investigated
in connection with $k$-inflation \cite{Arm1} and was
shown to produce exponential inflation.  Note that $X_0$ can be either a minimum
or a maximum of $F(X)$; the function $F(X)$ in equation (\ref{motion3})
does not correspond to a potential well.  It is also easy to see that this
solution is stable against small perturbations; taking $X = X_0 + \epsilon$,
and expanding equation (\ref{motion3}) out to order $\epsilon$, we obtain
\begin{equation}
\label{epsilon}
\dot \epsilon = - 3H\epsilon.
\end{equation}
In the analytic solution given by equation (\ref{solution}), this solution corresponds to taking
$k=0$ on the right-hand side.

Inserting this solution into equation (\ref{w}), we find that $w = -1$, so the
solution behaves like a cosmological constant.  The behavior of the sound speed,
however, is more interesting:  equation (\ref{cs}) gives $c_s^2 = 0$.  Thus,
this solution corresponds to an equation of state equivalent to a pure
cosmological constant, but with zero effective sound speed (as far as
perturbation growth is concerned).  As noted in Refs. \cite{Erickson,DeDeo,Bean},
the effect of $c_s^2 = 0$ is to suppress the integrated Sachs-Wolfe (ISW) effect at
large angular scales, since the quintessence component can cluster,
reducing the decay of the gravitational potental that causes the ISW
effect.

Although the final stable attractor solution in this case corresponds to a
field with $w = -1$ and $c_s^2 = 0$, consider how this state is
reached.  For $X$ near $X_0$, we assume that $F(X)$ can be expanded about its
minimum or maximum in the form:
\begin{equation}
\label{F(X)}
F(X) = F_0 + F_2(X-X_0)^2,
\end{equation}
where $F_0$ and $F_2$ are constants.
Substituting this form for $F(X)$ into equation (\ref{solution}) gives
\begin{equation}
\label{solution2}
4F_2^2(X - X_0)^2 X = k a^{-6},
\end{equation}
where $k$ is a constant.  Now suppose that $X$ is near the extremum of $F(X)$, so that
\begin{equation}
\label{smallX}
(X - X_0)/X_0 \ll 1.
\end{equation}
In this case, the solution in equation (\ref{solution2}) reduces to
\begin{equation}
\label{Xevol}
X = X_0 [1 + \epsilon_1 (a/a_1)^{-3}],
\end{equation}
where we have rewritten the solution in terms of the new constants
$\epsilon_1$ and $a_1$.  Note that we must have
\begin{equation}
\label{smallX2}
\epsilon_1 (a/a_1)^{-3} \ll 1,
\end{equation}
to satisfy equation
(\ref{smallX}).  This gives the physical meaning of $\epsilon_1$ and $a_1$:
our solution will be valid (i.e. condition \ref{smallX} will
be satisfied) for $a > a_1$, provided
that $\epsilon_1 \ll 1$.

The evolution of the density from equation (\ref{rho}) is
\begin{equation}
\rho = 4 F_2 X(X-X_0) - F_0 - F_2(X-X_0)^2.
\end{equation}
Substituting the evolution of $X$ from equation (\ref{Xevol}) into
this equation for the density gives
\begin{equation}
\label{rho2}
\rho = -F_0 + 4F_2 X_0^2 \epsilon_1 (a/a_1)^{-3} + 3F_2 X_0^2[\epsilon_1 (a/a_1)^{-3}]^2.
\end{equation}
Note that our condition in equation (\ref{smallX2}) implies that the third
term in equation (\ref{rho2}) is negligible compared to the second term, but it implies
nothing about the relative sizes of the first and second terms.  Hence, we have
\begin{equation}
\label{rho3}
\rho = -F_0 + 4F_2 X_0^2 \epsilon_1 (a/a_1)^{-3}.
\end{equation}
In order for this density to be positive at late times, we require $F_0 < 0$.

Equation (\ref{rho3}) implies that the $k$-essence energy density evolves as the sum
of a constant term and a term which evolves as non-relativistic matter.  Now
consider the effective sound speed for this fluid.  Substituting equation
(\ref{F(X)}) into equation (\ref{cs}), we get
\begin{equation}
c_s^2 = (X-X_0)/(3X - X_0),
\end{equation}
and using our solution for the evolution of $X$ (equation \ref{Xevol}), we get
\begin{equation}
\label{smallsound}
c_s^2 = \frac{1}{2}\epsilon_1 (a/a_1)^{-3}.
\end{equation}
Thus, equations (\ref{smallX2}) and (\ref{smallsound}) indicate that $c_s^2 \ll 1$ for the entire range of
validity of our solution.
Therefore, the $k$-essence fluid in this case behaves like a
low sound-speed fluid with a density which evolves like the sum of a
``dark matter" (DM) component with $\rho \propto a^{-3}$ and a ``dark energy" (DE)
component
with $\rho = constant$.  The only difference from a standard $\Lambda$CDM model is
that in this $k$-essence model, the dark energy component has $c_s^2 \ll 1$.

In practice,
the observational constraints on $\rho_{DM}$ and $\rho_{DE}$
sharply constrain the allowed parameters in this model.  The value of $a_1$ is
determined by the fact that the $k$-essence must begin to behave like dark matter
prior to
the epoch of equal matter and radiation.
Therefore, $a_1 < a_{eq}$, where $a_{eq}$ is the scale factor at the epoch of equal matter
and radiation, given by
$a_{eq}/a_0 = 3 \times 10^{-4}$, where $a_0$ is the value of the scale factor
today.  At the present time,
the component of $\rho$ corresponding to dark energy in equation (\ref{rho3})
must be roughly twice the component corresponding to dark matter, so
\begin{equation}
-F_0 = 8F_2 X_0^2 \epsilon_1 (a_0/a_1)^{-3}.
\end{equation}
Substituting $a_1 < a_{eq}$ into this equation, we get
\begin{eqnarray}
\label{limit}
\frac{-F_0} {F_2 X_0^2} &<& 8\epsilon_1(a_{eq}/a_0)^3, \\
&\ll& 2 \times 10^{-10}.
\end{eqnarray}
If $X_0$ is of order unity, this means that $-F_0/F_2 \ll 2 \times 10^{-10}$,
a somewhat unnatural fine-tuning.

When equation (\ref{smallX}) is not satisfied, i.e., for $a < a_1$, we can say
nothing about the evolution of $X$ and $\rho$, since we have assumed nothing about
$F(X)$, other than that it has a local minimum that can be expanded as a quadratic.
However, it is also interesting to consider what happens if $F(X)$ is
truly a quadratic function, as it is, for example, if one begins with equation
(\ref{Chibaeq}) and takes $K(\phi)$ and $L(\phi)$ to be constants.  In this
case, the solution to equation (\ref{solution}) when $X-X_0$ is not small compared to $X_0$ is
\begin{equation}
X \propto a^{-2},
\end{equation}
which gives, in equation (\ref{rho}),
\begin{equation}
\rho \propto a^{-4}.
\end{equation}
Thus, in this case, the $k$-essence evolves first like radiation, then like
dark matter, and finally like dark energy.  This resembles the phenomenological model
proposed by Cardone et al. \cite{Cardone}.  Note, however, that there is nothing desirable
about having a radiation-like behavior at early times.  The radiation content of the universe is
already well-explained in terms of the relic photon and neutrino backgrounds and, indeed, extra
relativistic energy density at early times is severely constrained by primordial
nucleosynthesis (see, e.g., Ref. \cite{Barger} and references therein).  Of course, an early
radiation-like behavior is not a generic prediction of this model, and, for the case of a
quadratic potential, the parameters can be tuned so that the contribution of the $k$-essence density
when it behaves like radiation can be made negligible.

\section{Discussion and Comparison with Other Models}

As noted in the previous section, the purely kinetic $k$-essence model, with a local extremum in
the Lagrangian $p = F(X)$, has an evolution which looks like the sum of a dark-matter component
(with $w=0$) and a dark energy component (with $w=-1$), but with a sound speed in both
components which satisfies $c_s^2 \ll 1$.  Thus, it is distinguishable from a standard
$\Lambda$CDM model in that the latter model has a purely non-clustering dark energy component.
As noted by DeDeo et al. \cite{DeDeo}, dark energy with an equation of state $w = -1$, but with
a negligible sound speed, is prevented from clustering by the Hubble expansion after the
dark energy component begins to dominate, but such a model predicts a very different
behavior for the large-angular scale CMB fluctuations \cite{DeDeo}.  In fact, such a model is
favored (albeit at a low significance level) by the CMB observations \cite{Bean}.

Other models proposed as unified dark matter can also be distinguished from this model.  In
the Chaplygin gas model \cite{Kamenshchik,Bilic}, the equation of state has the form
\begin{equation}
p = - \frac{A}{\rho},
\end{equation}
leading to a density evolution of the form
\begin{equation}
\rho = \sqrt{A + \frac{B}{a^6}}.
\end{equation}
In this model, the density interpolates between dark-matter-like evolution at early times and
dark-energy-like (i.e., constant density) evolution at late times, but in a different manner than in the model
presented here.  The Chaplygin gas model can be generalized \cite{Bento} to an equation of state
\begin{equation}
p = -\frac{A}{\rho^{\alpha}},
\end{equation}
which gives the density evolution
\begin{equation}
\rho = \left[A + \frac{B}{a^{3(1+\alpha)}}\right]^{1/(1+\alpha)}.
\end{equation}
Again, the density evolution in the generalized Chaplygin gas models changes from
$\rho \propto a^{-3}$ at early times to $\rho = constant$ at late times.  In the limit
where $\alpha \rightarrow 0$, the model approaches the behavior of the $k$-essence
model we have presented
here.

The sound speed in the generalized Chaplygin gas model is \cite{Makler}
\begin{equation}
c_s^2 = \alpha \left[1 + \frac{B/A}{a^{3(1+\alpha)}}\right]^{-1}.
\end{equation}
In the generalized Chaplygin gas model, the sound speed is small at early times (when $a$ is
small) and becomes larger at late times.  It has been argued that this behavior is
problematic for the model, since the large sound speed at late times leads to either blow-up
of the perturbations (for negative $\alpha$) or excessive damped oscillations (for positive
$\alpha$), with a resulting limit of $\alpha < 10^{-5}$ \cite{Sandvik}. (However,
nonlinear clustering may significantly alter the evolution of density perturbations in
this model \cite{Avelino}; see also the discussion in Refs. \cite{Finelli1,Finelli2}).

Our model, on the other hand, will not suffer from this particular problem, since the sound
speed can be made arbitrarily small during the epoch of structure formation by decreasing the
value of $\epsilon_1$ in equation (\ref{smallsound}).  Of course, this requires a decrease
in the ratio of parameters on the left-hand side of equation (\ref{limit}).  On the other hand,
nonlinear clustering will certainly modify this model.
In particular, the entire treatment in Sec. 2 assumes that spatial gradients can be neglected in
equation (\ref{grad}).  This will certainly not be the case when the clustering is nonlinear.

A related model, also based on purely kinetic $k$-essence, was proposed by
Chimento \cite{Chimento2}.  This model assumes a special form for $F(X)$, namely:
\begin{equation}
F(X) = A_1 \sqrt{X} - A_2 X^{\alpha},
\end{equation}
where $A_1$, $A_2$, and $\alpha$ are constants.  The resulting density evolution is of the form
\begin{equation}
\rho = \left [A + \frac{B}{a^3}\right]^{2\alpha/(2\alpha-1)}.
\end{equation}
This solution asymptotically approaches $\rho = constant$, for exactly the same reason that our
more general models do.  Further, the early evolution of this model resembles our own model in
the limit $\alpha \rightarrow \infty$.  There is, however, a fundamental difference between the
two models: the evolution of the Chimento model at early times is based on the large-$X$
evolution of the field, whereas the $a^{-3}$ behavior in our model is a transient effect, which
only shows up over a limited range in the evolution of $X$.

Finally, we mention another model proposed for unified dark matter, based on a tachyon field \cite{tachyon}.  This
model assumes a scale-dependent equation of state, so that the field behaves like pressureless dark
matter on small scales, and like smoothly-distributed dark energy on large scales.

The model presented here has both positive features and drawbacks.  On the positive
side, it arises from a very simple version of the $k$-essence model (indeed, perhaps the simplest
possible version) and can be achieved with very general assumptions about the behavior of
$F(X)$.  It produces unified dark matter with an equation of state identical to that of
ordinary dark matter plus a cosmological constant, but with the distinguishing feature that
the effective sound speed is small at all times.  This last characteristic avoids
the damping problems associated with the generalized Chaplygin gas models, and damps
CMB fluctuations on large scales.  On the negative side, since the
dark-matter-like evolution in this model is a transient phenomenon, the parameters of the model
must be fine-tuned in
a rather unnatural way.  Further, the model cannot (unlike the $k$-essence models of
Refs. \cite{Arm2,Arm3}) resolve the coincidence problem.  It is also far from clear whether
a model with $c_s^2 \ll 1$ in the dark-energy component is consistent with other
observations.

\acknowledgments

I thank C. Armendariz-Picon and M. Malquarti for helpful discussions.

\newcommand\AJ[3]{~Astron. J.{\bf ~#1}, #2~(#3)}
\newcommand\APJ[3]{~Astrophys. J.{\bf ~#1}, #2~ (#3)}
\newcommand\apjl[3]{~Astrophys. J. Lett. {\bf ~#1}, L#2~(#3)}
\newcommand\ass[3]{~Astrophys. Space Sci.{\bf ~#1}, #2~(#3)}
\newcommand\cqg[3]{~Class. Quant. Grav.{\bf ~#1}, #2~(#3)}
\newcommand\mnras[3]{~Mon. Not. R. Astron. Soc.{\bf ~#1}, #2~(#3)}
\newcommand\mpla[3]{~Mod. Phys. Lett. A{\bf ~#1}, #2~(#3)}
\newcommand\npb[3]{~Nucl. Phys. B{\bf ~#1}, #2~(#3)}
\newcommand\plb[3]{~Phys. Lett. B{\bf ~#1}, #2~(#3)}
\newcommand\pr[3]{~Phys. Rev.{\bf ~#1}, #2~(#3)}
\newcommand\PRL[3]{~Phys. Rev. Lett.{\bf ~#1}, #2~(#3)}
\newcommand\PRD[3]{~Phys. Rev. D{\bf ~#1}, #2~(#3)}
\newcommand\prog[3]{~Prog. Theor. Phys.{\bf ~#1}, #2~(#3)}
\newcommand\RMP[3]{~Rev. Mod. Phys.{\bf ~#1}, #2~(#3)}

\end{document}